\documentclass{WileyMSP-template}
\usepackage{xcolor}

\begin{document}

\pagestyle{fancy}
\rhead{\includegraphics[width=2.5cm]{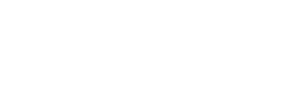}}

\title{Bidirectional motion of antiferromagnetic skyrmions driven by competing spin torques}

\maketitle


\author{Laichuan Shen}
\author{Wang Kang}
\author{Xichao Zhang}
\author{Qiuping Huang}
\author{Yalin Lu}
\author{Zhifeng Zhu*}
\author{Yan Zhou*}



\begin{affiliations}
Laichuan Shen, Yan Zhou\\
School of Science and Engineering, The Chinese University of Hong Kong, Shenzhen, Guangdong 518172, China\\
Email Address: zhouyan@cuhk.edu.cn

Wang Kang\\
School of Integrated Circuit Science and Engineering, Beihang University, Beijing, 100191, China\\
State Key Laboratory of Spintronics, Hangzhou International Innovation Institute, Beihang University, Hangzhou, 311115, China

Xichao Zhang\\
Department of Electronic and Computer Engineering, The Hong Kong University of Science and Technology, Clear Water Bay, Kowloon, Hong Kong, China\\
IAS Center for Quantum Matter, The Hong Kong University of Science and Technology, Hong Kong, China

Qiuping Huang, Yalin Lu\\
Hefei National Laboratory, University of Science and Technology of China, Hefei 230088, China

Zhifeng Zhu\\
School of Information Science and Technology, ShanghaiTech University, Shanghai 201210, China\\
Shanghai Engineering Research Center of Energy Efficient and Custom AI IC, Shanghai 201210, China\\
Email Address: zhuzhf@shanghaitech.edu.cn

\end{affiliations}



\begin{abstract}

 Antiferromagnetic skyrmions are swirling topological spin textures with rich dynamics and intriguing transport properties, yet their bidirectional dynamics remain largely unexplored. Here, we investigate the dynamics of antiferromagnetic skyrmions driven by current-induced spin-transfer and spin-orbit torques. We computationally demonstrate that antiferromagnetic skyrmions moving in one direction at low current densities can reverse their motion direction when the driving current is above a threshold. Based on the Thiele approach analysis, we show that this bidirectional motion originates from a change in the relative strengths of two effective forces arising from spin-transfer and spin-orbit torques. Furthermore, exploiting this bidirectional motion on a single racetrack, we design programmable logic gates. Our results not only uncover a hidden mechanism for bidirectional skyrmion motion but also facilitate the development of antiferromagnet-based logic devices.

\end{abstract}


\section{Introduction}

Directional control of particle-like entities is central to diverse fields, from soft matter and biological locomotion to condensed matter physics, and has been widely investigated for various particles and quasiparticles, including particles in ratchet systems~\cite{hanggi_artificial_2009,mateos_chaotic_2000}, solitons in liquid crystals~\cite{zhang_autonomous_2021}, Janus particles~\cite{vutukuri_light-switchable_2020}, bacteria~\cite{son_bacteria_2013}, superconducting vortices~\cite{de_souza_silva_controlled_2006,villegas_superconducting_2003,reichhardt_depinning_2017}, Yukawa particles in one-dimensional channels~\cite{reichhardt_positive_2011}, and magnetic solitons~\cite{wang_magnon-driven_2015,oh_bidirectional_2019,zhang_skyrmion_2017,hirata_vanishing_2019,liang_bidirectional_2023,shen_anomalous_2025,reichhardt_statics_2022}. One remarkable manifestation of such control is transport-direction reversal, in which the drift velocity of particles or quasiparticles changes sign under modified driving conditions. For instance, Janus particles with different photocatalysts on their two sides exhibit bidirectional motion when illuminated with light of different wavelengths~\cite{vutukuri_light-switchable_2020}.

Similar bidirectional phenomena have recently attracted increasing attention in magnetic systems. In ferrimagnets, bidirectional motion of domain walls driven by spin waves (magnons) has been reported, with the motion direction depending on the sign of net spin density~\cite{oh_bidirectional_2019}. Bidirectional magnon-driven motion was also predicted for ferromagnetic bimerons, which can be either pushed away from or pulled toward the wave source depending on the spin-wave frequency~\cite{liang_bidirectional_2023}. In addition, skyrmions in frustrated ferromagnets have been found to exhibit transport reversal by precisely controlling their helicity with current~\cite{zhang_skyrmion_2017,lin_ginzburg-landau_2016}. Recent work has shown that the motion direction of skyrmion pairs in a synthetic antiferromagnet can occasionally reverse due to thermal fluctuations~\cite{de_souza_silva_emergent_2025}. These results indicate that magnetic textures provide a versatile platform for realizing tunable bidirectional transport, owing to their particle-like nature and rich interactions with external excitations.

In particular, magnetic skyrmions are topologically nontrivial magnetic textures with promising potential for spintronic applications, including racetrack memories, logic devices, and nano-oscillators~\cite{bogdanov_thermodynamically_1989,rosler_spontaneous_2006,muhlbauer_skyrmion_2009,nagaosa_topological_2013,zhou_magnetic_2019,gobel_beyond_2021,fert_magnetic_2017,everschor-sitte_perspective_2018,zhang_skyrmion-electronics_2020,back_2020_2020,kang_skyrmion-electronics_2016,zhou_condensed_2025}. Their dynamics, driven by current-induced spin-transfer and spin-orbit torques, have been extensively studied in ferromagnets~\cite{sampaio_nucleation_2013,masell_spin-transfer_2020}, ferrimagnets~\cite{woo_current-driven_2018}, and antiferromagnets~\cite{zhang_antiferromagnetic_2016,barker_static_2016,jin_dynamics_2016,shen_current-induced_2019,pham_fast_2024}. Antiferromagnetic skyrmions are particularly attractive due to their high mobility and the absence of the skyrmion Hall effect~\cite{jiang_direct_2017,litzius_skyrmion_2017}, making them promising information carriers for high-density and high-speed spintronic devices. Nevertheless, most previous studies have focused on unidirectional current-driven motion, while the possibility of reversing the motion direction by tuning the current magnitude remains largely unexplored.

In this work, we demonstrate that antiferromagnetic skyrmions driven by current-induced spin-transfer and spin-orbit torques can exhibit bidirectional motion: at low current, the skyrmion moves along the current direction, whereas above a threshold current its motion direction reverses. This behavior contrasts sharply with the conventional scenario, where the skyrmion velocity simply increases in magnitude while maintaining its direction as the current increases. We attribute this phenomenon to the reversal of the net force from the two competing spin torques, induced by the change in skyrmion size. Our results reveal a distinct mechanism for bidirectional skyrmion transport and provide a useful strategy for designing programmable skyrmion logic devices.

\section{Results and Discussion}

\subsection{Theoretical model}

\begin{figure}
   \centering
  \includegraphics[width=0.9\linewidth]{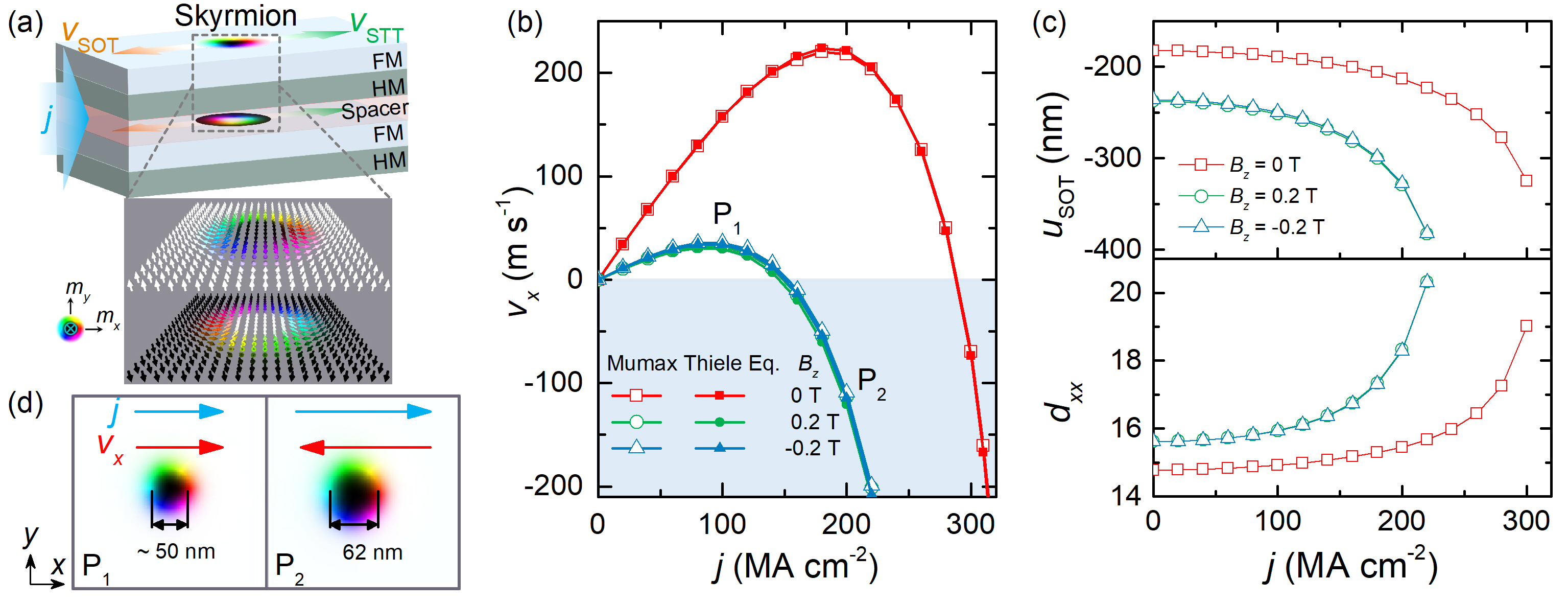}
  \caption{(a) Top panel: Sketch of the calculation model, where the two-sublattice antiferromagnet is composed of stacked ferromagnet (FM)/heavy metal (HM) layers. The spacer is used to induce interlayer coupling. The current $j$ is simultaneously injected into the FM and HM layers. The yellow and green arrows represent the velocities of skyrmions driven by SOT and STT, respectively. Bottom panel: Magnetic texture of an antiferromagnetic skyrmion in a two-sublattice model, with colored areas reflecting the in-plane component of magnetization and black (white) areas indicating the magnetizations along the $-z$ ($+z$) direction.
 (b) Skyrmion velocity $v_{x}$ as a function of the current density $j$ with different values of magnetic field $B_{z}$. Hollow and solid symbols represent the results obtained from micromagnetic simulations and Thiele equation, respectively. The colored background indicates $v_{x} <0$. P$_1$ and P$_2$ are two representative points indicating that $v_{x} > 0$ at a small current ($j = 100$~MA~cm$^{-2}$) and $v_{x} < 0$ at a large current ($j = 200$~MA~cm$^{-2}$), with $B_{z}=0.2$~T.
  (c) Parameters $u_{\rm{SOT}}$ and $d_{xx}$ involved in the skyrmion velocity formula.
  (d) Snapshots of the dynamically stable skyrmion structure at simulation time 2~ns for the conditions indicated by points P$_1$ and P$_2$ in panel (b). The cyan and red arrows represent the applied current and skyrmion velocity, respectively.
  }
  \label{FIG1}
\end{figure}

We consider a two-sublattice antiferromagnet with reduced magnetizations $\mathbf{m}^{\rm{t}}$ and $\mathbf{m}^{\rm{b}}$, where the intralayer exchange interaction between adjacent magnetizations is ferromagnetic (with exchange coefficients $A_{\rm{ex}}>0$), while the interlayer magnetizations are coupled through an antiferromagnetic exchange interaction ($A_{\rm{int}} <0$). To stabilize antiferromagnetic skyrmions, heavy metal layers are introduced into the model to induce an interfacial Dzyaloshinskii-Moriya interaction~\cite{dzyaloshinsky_thermodynamic_1958,moriya_anisotropic_1960}, as illustrated in Figure~\ref{FIG1}(a). The time evolution of magnetizations is computed by utilizing the micromagnetic simulation package Mumax3~\cite{vansteenkiste_design_2014} to solve the Landau-Lifshitz-Gilbert (LLG) equation~\cite{gilbert_phenomenological_2004} incorporating current-induced spin-transfer torque (STT) and spin-orbit torque (SOT) (see Supporting Information for simulation details). In this work, the two current-induced spin torques, described as~\cite{sampaio_nucleation_2013,finocchio_magnetic_2016,manchon_current-induced_2019}
\begin{equation}
	\begin{array}{cc}
\mathbf{T}_{\mathrm{STT}}^{\tau}=-\mathit{u}\mathit{\partial_{x}}\mathbf{m}^{\mathrm{\tau}}+\beta\mathit{u}\mathbf{m}^{\mathrm{\tau}}\times\mathit{\partial_{x}}\mathbf{m}^{\mathrm{\tau}} \\
\mathbf{T}_{\mathrm{SOT}}^{\tau}=\gamma H_{j}\left(\mathbf{m}^{\mathrm{\tau}}\times\mathbf{p}\right)\times\mathbf{m}^{\mathrm{\tau}},
	\label{eq:1}
\end{array}
\end{equation}
drive the skyrmion to move in opposite directions [Figure~\ref{FIG1}(a)], where $\tau =$ t, b.
The charge drift velocity $u$ is defined as $u = gu_{\mathrm{B}}Pj/\left[2eM_{\mathit{\mathrm{s}}}\left(1+\beta^{2}\right)\right]$ with $g$ being the $g$-factor, $u_{\mathrm{B}}$ the Bohr magneton, $P$ the spin polarization efficiency, $j$ the current density, $e$ the elementary charge, $M_{\mathrm{s}}$ the saturation magnetization and $\beta$ the nonadiabatic parameter.
$\gamma$ denotes the gyromagnetic ratio and $\mathbf{p}$ is the polarization vector. $H_{j}$ is the strength of SOT, expressed as $H_{j}=j\hbar \theta_{\rm{SH}}/(2\mu_{0}\mathit{e t_{z}\mathit{M}_{\mathrm{s}}})$ with $\hbar$ being the reduced Planck constant, $\theta_{\rm{SH}}$ the spin Hall angle, $\mu_0$ the vacuum permeability constant, and $t_{z}$ the sublayer thickness.

We utilize the phenomenological Thiele equation~\cite{thiele_steady-state_1973} to analyze the current-induced dynamics of antiferromagnetic skyrmions, which is written as~\cite{shen_current-induced_2019,vakili_spin-transfer_2025,shen_current-induced_2020}: 
\begin{equation}
m_{\mathrm{eff}}\ddot{r}_{\mathrm{c}}=-\alpha L\dot{r}_{\mathrm{c}}+F_{\mathrm{STT}}+F\mathrm{_{SOT}},
		\label{eq:2}
		\end{equation}
where $m_{\mathrm{eff}}=-\mu_{0}M_{\rm{s}}t_{z} d_{xx}/(2\gamma^{2}\mathit{H}_{\mathit{\mathrm{int}}})$ is the effective mass, with the interlayer exchange field $H_{\rm{int}}=\mathit{\mathrm{2}A_{\mathrm{int}}} /(\mu_{\mathrm{0}}\mathit{M}_{\mathrm{s}}\mathit{t}_{\mathit{z}}^{\mathrm{2}})$ and dimensionless parameter $d_{xx}=\int\mathit{\partial_{x}}\mathbf{n}\cdot\mathit{\partial_{x}}\mathbf{n}dxdy$. $\mathbf{n} = (\mathbf{m}^{\rm{t}}-\mathbf{m}^{\rm{b}})/2$ is the N{\'e}el vector.
$r_{\rm{c}}$ stands for the skyrmion location, given by $r_{\rm{c}}=\int x\rho_{\mathrm{s}}dxdy/\int\rho_{\mathrm{s}}dxdy$ with topological density $\rho_{\mathrm{s}}=1/(4\pi)\mathbf{n}\cdot\left(\mathit{\partial_{x}}\mathbf{n}\times\mathit{\partial_{y}}\mathbf{n}\right)$~\cite{komineas_skyrmion_2015}.
 The first term on the right-hand side of the Thiele equation is the dissipation force with the magnetic damping $\alpha$ and dissipation coefficient $L=\mu_{0}M_{\rm{s}}t_{z}d_{xx} /\gamma$. The remaining two terms, $F_{\mathrm{STT}}=\beta u L$ and $F\mathrm{_{SOT}} = -\mu_{0}M_{\rm{s}}H_{j}t_{z} u_{\rm{SOT}}$, are the current-induced forces from STT and SOT, respectively. Here, $u_{\rm{SOT}} = \int\left(\mathbf{n}\times\mathbf{p}\right)\cdot\partial_{x}\mathbf{n}dxdy$.

From the Thiele equation, the skyrmion velocity can be derived as
\begin{equation}
v_{x}=\dot{r}_{\mathrm{c}}=\frac{\beta\mathit{u}}{\alpha}-\frac{\gamma H_{j}u_{\mathrm{SOT}}}{\alpha d_{xx}}.
\label{eq:3}
\end{equation}
The velocity formula indicates that the STT-induced velocity $\beta\mathit{u}/\alpha$ is independent of the skyrmion structure, while the parameters $u_{\mathrm{SOT}}$ and $d_{xx}$ in the SOT-induced velocity $-\gamma H_{j}u_{\mathrm{SOT}}/(\alpha d_{xx})$ are determined by the magnetic structure. This inspires us to modulate the SOT-induced velocity by adjusting the magnetic structure, causing a change in the relative magnitudes of the SOT- and STT-induced velocities, thereby achieving skyrmion velocity reversal. Equation~(\ref{eq:3}) also suggests that $\beta\mathit{u} \sim \gamma H_{j}u_{\mathrm{SOT}}/d_{xx}$ is desirable for observing velocity reversal.
In the actual implementation of the experiment, two key conditions are involved. The first condition is to realize competing STT and SOT, which can be achieved by selecting appropriate heavy metals. For example, Pt and Ta possess opposite spin Hall angles~\cite{hahn_comparative_2013}, leading to opposing SOT-induced driving forces, whereas the direction of the STT-induced force remains largely unaffected. The second condition requires the STT- and SOT-induced velocities to be of comparable magnitude. This can be achieved by tuning the skyrmion structure (e.g., by applying a magnetic field) and the magnetic layer thickness to modulate the SOT-induced velocity, as indicated by Eq.~(\ref{eq:3}).

\subsection{Bidirectional motion of antiferromagnetic skyrmions}

Generally, an electric current drives an antiferromagnetic skyrmion to expand~\cite{barker_static_2016,jin_dynamics_2016}, so that the current may act as a lever to control the skyrmion structure and hence its velocity. To verify the validity of this prediction, we perform micromagnetic simulations and plot the skyrmion velocity $v_{x}$ as a function of current density $j$ in Figure~\ref{FIG1}(b) with different values of out-of-plane magnetic field $B_{z}$. The results from the micromagnetic simulations agree well with those from Eq.~(\ref{eq:3}) which uses the numerical values of $u_{\rm{SOT}}$ and $d_{xx}$ in Figure~\ref{FIG1}(c). As shown in Figure~\ref{FIG1}(b), the skyrmion velocity exhibits a non-monotonic variation with increasing current, changing from positive to negative and indicating a reversal in the motion direction. When an out-of-plane magnetic field $B_{z}$ is applied, we still observe a velocity reversal, but the critical current that triggers the reversal decreases, almost independent of the sign of $B_{z}$. Thus, we demonstrate not only the bidirectional motion of the antiferromagnetic skyrmion by varying the current density but also its tunability by a magnetic field. Note that the critical current cannot be effectively regulated by the interlayer exchange coefficient and magnetic damping (see Figure S1). We further present representative skyrmion structures in Figure~\ref{FIG1}(d) for the cases of $v_{x}>0$ and $v_{x}<0$, confirming that the velocity reversal is caused by a change in the skyrmion size. 
Unlike the stochastic, thermally induced direction reversals of self-propelled skyrmion pairs in synthetic antiferromagnets~\cite{de_souza_silva_emergent_2025}, the reversal observed here can be controlled deterministically.

\begin{figure}[t]
   \centering
  \includegraphics[width=0.9\linewidth]{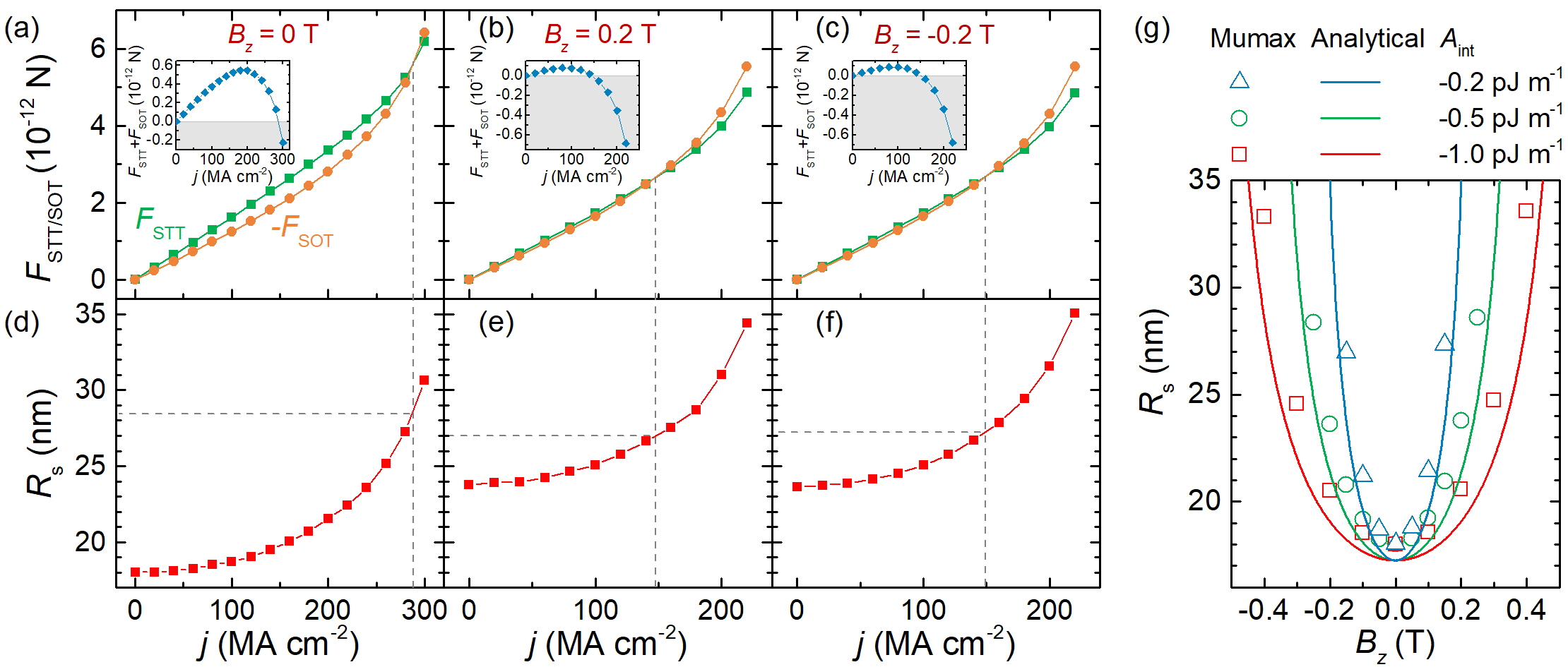}
  \caption{(a)-(c) Calculated forces induced by STT ($F_{\rm{STT}}$) and SOT ($F_{\rm{SOT}}$) as a function of current density $j$ for different values of $B_{z}$. The vertical dashed line indicates the current at which the strengths of $F_{\rm{STT}}$ and $F_{\rm{SOT}}$ are reversed. The insets show the net force $(F_{\rm{STT}}+F_{\rm{SOT}})$.
  (d)-(f) Skyrmion radius $R_{\rm{s}}$ as a function of the current $j$, where $R_{\rm{s}}$ is defined as the radius of the contour with $n_{z} =0$.
  (g) The dependence of the skyrmion radius $R_{\rm{s}}$ on the magnetic field $B_{z}$, with different values of the interlayer coupling strength $A_{\rm{int}}$. The symbols are the simulation results with $j=0$; the curves represent the analytical results.
  }
  \label{FIG2}
\end{figure}

To gain more insights into the bidirectional motion of antiferromagnetic skyrmions, we compute two current-induced forces $F_{\mathrm{STT}}$ and $F\mathrm{_{SOT}}$ as a function of the current density $j$ with different values of $B_{z}$, as shown in Figure~\ref{FIG2}(a)-(c). The directions of the forces induced by the SOT and STT are opposite, and their magnitudes exhibit distinct growth rates with increasing current, which can be understood from the fact that the two forces have different expressions [see Eq.~(\ref{eq:2})]. As a result, the relative strengths of two forces change above a certain current density, giving rise to the reversal of the net force ($F_{\mathrm{SOT}}+F\mathrm{_{STT}}$) and hence the skyrmion motion direction. 
To quantify the change in skyrmion size, based on space-dependent magnetizations output from the micromagnetic simulations, we extract skyrmion radius $R_{\rm{s}}$ defined as the radius of the contour with out-of-plane N{\'e}el vector $n_{z} =0$ and present $R_{\rm{s}}$ in Figure~\ref{FIG2}(d)-(f). As seen, the critical $R_{\rm{s}}$ at which the reversal of the net force occurs is around 27-28 nm for the parameters we used. On the other hand, we can attain the critical $R_{\rm{s}}$ from Eq.~(\ref{eq:3}) with $v_{x}=0$. Specifically, substituting the approximate expressions for $u_{\rm{SOT}}$ and $d_{xx}$ into Eq.~(\ref{eq:3})~\cite{shen_spin_2019}
\begin{equation}
u_{\rm{SOT}} \approx -\pi^{2}R_{\mathrm{s}},
\label{eq:4}
\end{equation}
\begin{equation}
d_{xx} \approx 2\pi\left[\left(R_{\mathrm{s}}^{2}/l_{\mathrm{DW}}^{2}+1\right)^{1/2}+\left(R_{\mathrm{s}}^{2}/l_{\mathrm{DW}}^{2}+1\right)^{-1/2}\right],
\label{eq:5}
\end{equation}
the critical $R_{\rm{s}}$ of $\sim 26$~nm is given, which is close to the numerical results. Here, $l_{\mathrm{DW}} = \sqrt{A_{\rm{ex}}/K_{\rm{eff}}}$ describes the domain wall width, with $K_{\rm{eff}} = K-\mu_{0} M_{\rm{s}}^{2}/2$ and anisotropy constant $K$. 
As illustrated in Figure~\ref{FIG2}(d)-(f), skyrmion expansion persists even under a zero net force, which could be attributed to the presence of the generalized forces $Q_{R_{\mathrm{s}}}$ and $Q_{l}$. Specifically, the change in the skyrmion radius satisfies $[R_{\mathrm{s}}(j)-R_{\mathrm{s}}(0)]\propto c_{1} Q_{R_{\mathrm{s}}}+c_{2} Q_{l}$ with the coefficients $c_{1(2)}$ (see Supporting Information for details).

Figure~\ref{FIG2}(d)-(f) indicate that the magnetic field can adjust the size of skyrmion and thus its velocity. To reveal the relationship between the magnetic field and skyrmion size, we derive the contribution of the magnetic field to the effective field under the continuous approximation (see Supporting Information), suggesting that a positive or negative magnetic field reduces the magnetic anisotropy:
\begin{equation}
K\rightarrow K+\frac{M_{\mathrm{s}}B_{z}^{2}}{4\mu_{0}H_{\mathrm{int}}},
\label{eq:6}
\end{equation}
where $H_{\mathrm{int}}<0$. Such a reduction in magnetic anisotropy leads to an increase in the skyrmion radius~\cite{wang_theory_2018}
\begin{equation}
R_{\mathrm{s}} \approx \pi D\sqrt{\frac{A_{\rm{ex}}}{16A_{\rm{ex}}K_{\rm{eff}}^{2}-\pi^{2}D^{2}K_{\rm{eff}}}},
\label{eq:7}
\end{equation}
where $D$ is the strength of Dzyaloshinskii-Moriya interaction. To verify the analytical result, we simulate the configurations of a static skyrmion under different magnetic fields $B_{z}$ and interlayer exchange coefficients $A_{\rm{int}}$, and find that the numerical $R_{\rm{s}}$ is consistent with the result given by Eq.~(\ref{eq:6})-(\ref{eq:7}), as shown in Figure~\ref{FIG2}(g).

\subsection{Programmable skyrmion-based logic gates}

\begin{figure}[t]
   \centering
  \includegraphics[width=0.9\linewidth]{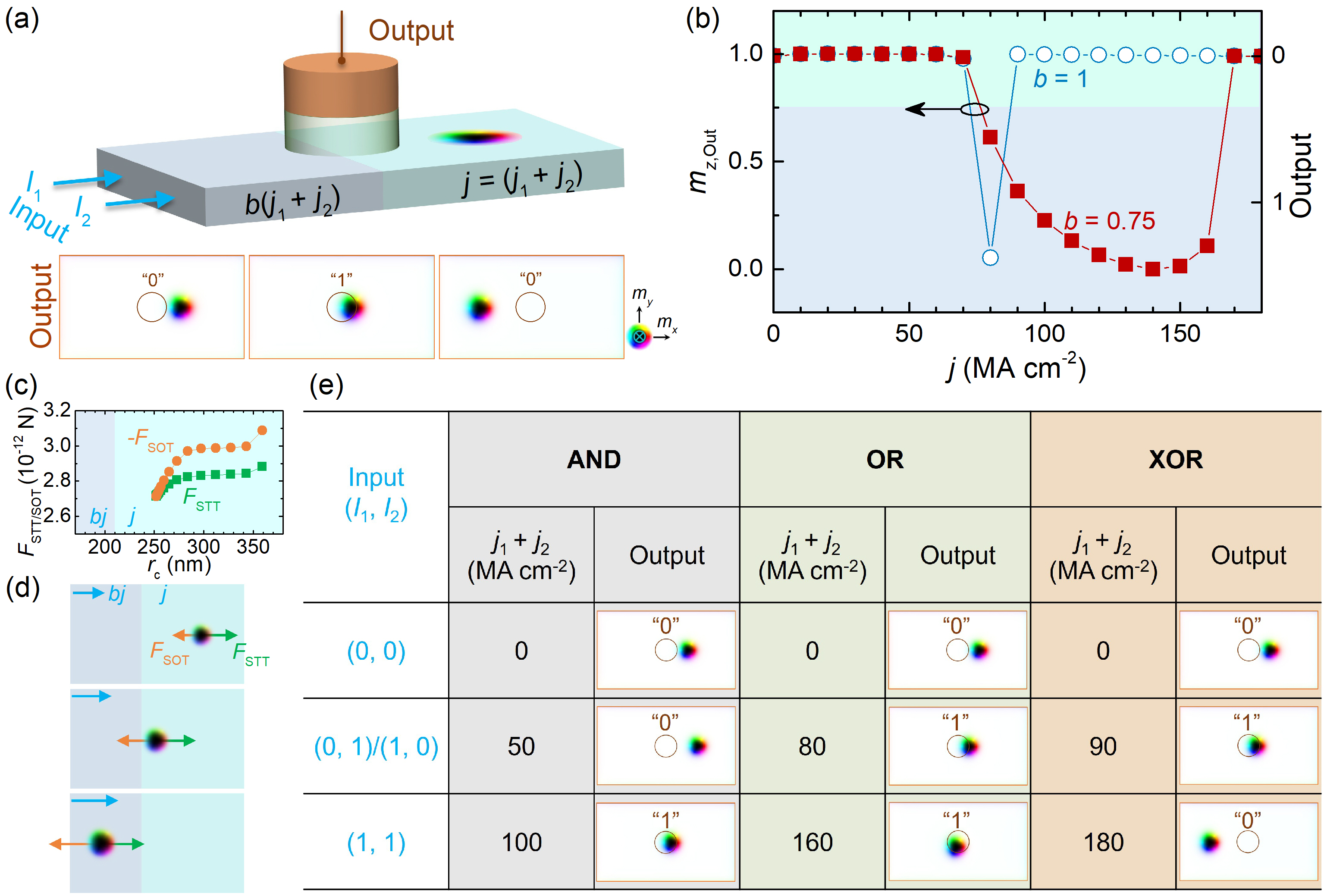}
  \caption{(a) Schematic of the skyrmion-based logic device. Two currents $I_1$ and $I_2$ (with current densities $j_1$ and $j_2$) are designed as logic inputs; logic input ``1'' (``0'') corresponds to (not) applying current to the device. The logic outputs ``1'' and ``0'' are defined as the presence and absence of the skyrmion underneath the middle detector, respectively (bottom panel). Here, $b$ represents the ratio of the current density on the left side of racetrack to that on the right side. 
  (b) Magnetization $m_{z,\mathrm{Out}}$ of the sublayer underneath the middle detector for different current densities $j$ with two values of $b=1$ and $0.75$. The decrease in $m_{z,\mathrm{Out}}$ from $\sim +1$ is due to the skyrmion entering the detection region; the corresponding logic outputs are marked with light blue and cyan backgrounds.
  (c) Calculated forces induced by SOT and STT as a function of skyrmion position $r_{\rm{c}}$. In this calculation, we take $j=150$~MA~cm$^{-2}$ and $b=0.75$. 
  (d) Illustration of the magnitude of SOT- and STT-induced forces for different current intensities. 
  (e) Demonstration of AND, OR and XOR logic operations.
  In the simulations, the model size is $500 \times 280 \times 3$~nm$^{3}$, the diameter of circular detector is 80~nm, the simulation time is 25~ns, $B_{z} = 0.23$~T, and $\alpha = 0.05$.}
  \label{FIG3}
\end{figure}

We further exploit the intriguing bidirectional dynamics of antiferromagnetic skyrmions to design programmable logic gates, which could serve as essential building blocks in future antiferromagnetic spintronic devices. As shown in Figure~\ref{FIG3}(a), our logic device mainly consists of a nanoscale racetrack and a detector above it, where two currents $I_1$ and $I_2$ injected into the racetrack correspond to two logic inputs, while logic outputs are defined as the presence and absence of the skyrmion underneath the detector [see bottom panel of Figure~\ref{FIG3}(a)]. The detector captures skyrmion-induced magnetization variations via magnetoresistance effects and converts them into electrical signals~\cite{kang_skyrmion-electronics_2016}. For small net current $I_1+I_2$ with density $j_1+j_2$, the skyrmion tends to be located on the right side of racetrack, while it moves to the left side for large currents, as indicated by Figure~\ref{FIG1}(b). These scenarios yield the logic output ``0''. When a suitable current is applied, the skyrmion enters the middle detection region, causing a decrease in the magnetizations of that region, corresponding to the logic output ``1'', as shown in Figure~\ref{FIG3}(b). However, the current that can produce a logic output ``1'' has a narrow range for a spatially uniform current distribution [see the hollow circle symbols in Figure~\ref{FIG3}(b)], which is detrimental to the implementation of OR logic gate, since it requires both small and large currents to give a logic output ``1'' (as discussed later). 
This problem can be solved by reducing the current density on the left side of racetrack with reduction factor $b$. Note that the value of $b$ can be adjusted by changing the cross-sectional area of a portion of racetrack, more specifically, changing the racetrack width~\cite{zhou_reversible_2014}. As indicated by the solid square symbols in Figure~\ref{FIG3}(b) for $b=0.75$, the lower limit of current density for logic output ``1'', determined by the velocity-reversal current discussed above, remains unchanged. By contrast, the upper limit of current density is significantly increased, meaning that the skyrmion can be blocked in the middle region of racetrack even with a relatively large current. This phenomenon is similar to a skyrmion encountering a barrier, but unlike common methods that modify magnetic parameters or film thickness to construct a barrier~\cite{zhao_realization_2024,sisodia_robust_2022,juge_helium_2021,jin_high-frequency_2020}, here we only change the local current intensity. To reveal the physics behind this phenomenon, we perform a simulation with a skyrmion gradually moving from a high-current region to a low-current region, and calculate the SOT- and STT-induced driving forces ($F_{\mathrm{SOT}}$ and $F_{\mathrm{STT}}$) at different skyrmion locations ($r_{\mathrm{c}}$). As shown in Figure~\ref{FIG3}(c), there is a finite net driving force ($F_{\mathrm{SOT}}+F_{\mathrm{STT}}$) for the skyrmion in the high-current region, while it is close to zero when the skyrmion approaches the low-current region. The weakening of the driving forces can be attributed to changes in the skyrmion structure. Figure~\ref{FIG3}(d) illustrates the position and the driving forces of the skyrmion under different current intensities.  

\begin{figure}
   \centering
  \includegraphics[width=0.5\linewidth]{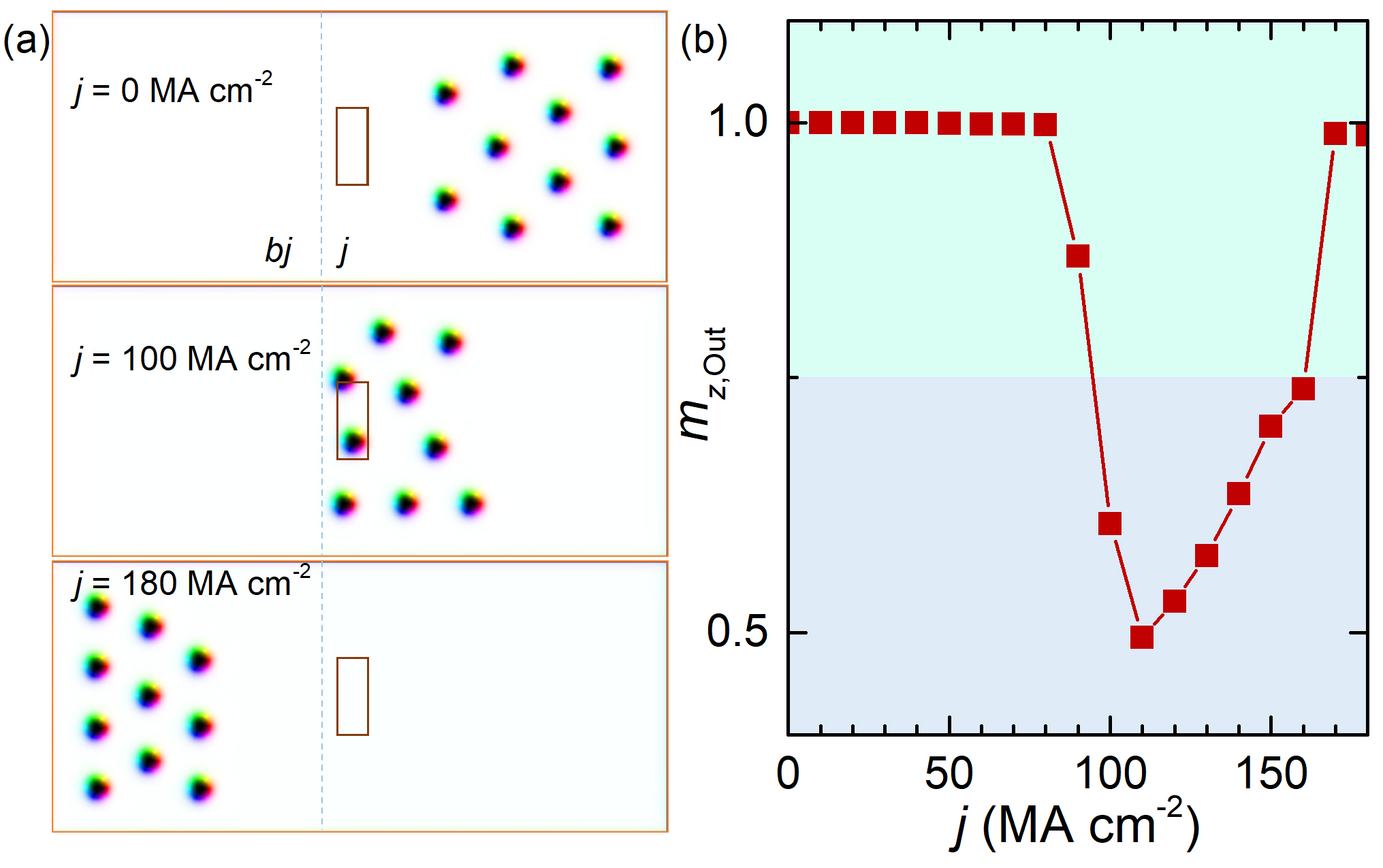}
  \caption{
  (a) Snapshots of magnetic structures under different currents. The box in the middle indicates the detection area.
  (b) Magnetization $m_{z,\mathrm{Out}}$ of the sublayer underneath the detector for different current densities $j$.
  In the simulations, the model size is $1600 \times 700 \times 3$~nm$^{3}$, the size of square detector is $80 \times 200$~nm$^{2}$, the simulation time is 60~ns, $b=0.75$, $B_{z} = 0.23$~T, and $\alpha = 0.05$.
  }
  \label{FIG4}
\end{figure}

With the relationship between the logic inputs and outputs indicated by solid square symbols in Figure~\ref{FIG3}(b), we can perform various logic operations by adjusting the intensity of input currents. Specifically, for AND logic gate, the logic inputs (0, 0)/(0, 1)/(1, 0) yield the logic output ``0'', so that the total input current density $j = (0+0)/(0+j_{2})/(j_{1}+0)$ should not exceed $80$~MA~cm$^{-2}$. Here we consider $j_{1}=j_{2}$.  
On the other hand, the logic input (1, 1) gives a logic output ``1'', which means that the total current density $j = (j_{1}+j_{2})$ should be in the range of [80, 160]~MA~cm$^{-2}$. As shown in Figure~\ref{FIG3}(e), taking $j_{1}=j_{2}=50$~MA~cm$^{-2}$ as an example, the AND logic operation is executed.
The OR logic gate requires the logic inputs (0, 1)/(1, 0)/(1, 1) to produce the output ``1''; therefore, ($0+j_2$)/($j_1+0$)/($j_1+j_2$) must fall within this range [80, 160]~MA~cm$^{-2}$. Apparently, $j_{1}=j_{2}=80$~MA~cm$^{-2}$ satisfies the requirement. For XOR logic gate, the output is ``1'' for inputs (0, 1)/(1, 0) and ``0'' for (0, 0)/(1, 1), which indicates that $j_1$ and $j_2$ are within the range [80, 160]~MA~cm$^{-2}$ and their sum exceeds the upper limit of this range. The above analysis indicates that the operating current-density window of the logic gates is determined by the lower and upper limits of the range, which are governed by the velocity-reversal current and the reduction factor, respectively. Such lower and upper limits tend to decrease when thermal fluctuations are taken into account (see Figure S2). Once the definition of the logic output is reversed, the logic gates AND, OR, and XOR are transformed into NAND, NOR, and XNOR, respectively~\cite{shen_programmable_2023}.

When multiple skyrmions are placed in the racetrack and driven by the currents [see Figure~\ref{FIG4}(a)], they exhibit similar states to those in Figure~\ref{FIG3}(a) where only a single skyrmion is considered, indicating high scalability for our logic devices. Specifically, for a logic device with multiple skyrmions, one can obtain input-output relationships [Figure~\ref{FIG4}(b)] similar to those shown in Figure~\ref{FIG3}(b), thereby enabling the execution of relevant logic operations. Such a multi-skyrmion logic device does not involve deterministic generation and precise detection of a single skyrmion~\cite{zhang_nanofluidic_2025}, thus facilitating the device implementation.
Although the current density used here is experimentally achievable~\cite{pham_fast_2024}, it could be higher than that of common skyrmion logic devices~\cite{yu_skyrmions-based_2022,yan_skyrmion-based_2021}. The operating current can actually be optimized, for example, by selecting a suitable magnetic field/reduction factor (see Figure S2) and adopting appropriate materials with easily tunable skyrmion sizes.
Compared to common skyrmion logic devices, the proposed logic gates—although entailing a trade-off in energy consumption due to the high current requirements (see Table S1)—do not necessitate precise, nanoscale control over the properties of the magnetic layer~\cite{shen_programmable_2023}, which greatly facilitates their practical implementation.
Our results can be directly transferred to other types of antiferromagnets~\cite{barker_static_2016,darwin_biasengineered_2026}, and the reversal of skyrmion velocity is also observed in an altermagnet~\cite{vakili_spin-transfer_2025,jin_skyrmion_2024} (see Figure S3).
We also note that the material inhomogeneity and the asymmetric current distribution within the magnetic and heavy-metal layers do not affect our main conclusions (see Figure S4, S6 and S7).

\section{Conclusion}

In summary, we have theoretically investigated the dynamics of antiferromagnetic skyrmions driven by current-induced spin-transfer and spin-orbit torques. We have demonstrated that the velocity of skyrmions under these two driving mechanisms can be effectively tuned by current-controlled modulation of the skyrmion size. The simulations have revealed bidirectional skyrmion motion and its tunability by an external magnetic field, which are in line with the Thiele-equation analysis. Furthermore, we have proposed a programmable logic device based on the motion of antiferromagnetic skyrmions in a single racetrack. Our findings provide insights into the bidirectional motion of skyrmions and may contribute to the development of antiferromagnetic skyrmion-based spintronic devices.

\medskip
\textbf{Supporting Information} \par 
Additional supporting information can be found online in the Supporting Information section.

Supporting File: XXX.PDF

\medskip
\textbf{Acknowledgements} \par 
Y.Z. acknowledges the support from the National Natural Science Foundation of China (Grant No. 12374123 and No. 11974298).
Q.H. and Y.L. were supported by the HFNL Self-Deployed Project (ZB2025020100).
X.Z. acknowledges support by the Grants-in-Aid for Scientific Research from JSPS KAKENHI (Grant No. JP25K17939 and No. JP20F20363).
L.S. acknowledges the support from the National Natural Science Foundation of China (Grant No. 12504151).

\medskip
\textbf{Data Availability Statement}  \par
The data supporting the findings of this study are available within this article and its Supporting Information. Additional data that support the findings of this study are available from the corresponding author on reasonable requests. The micromagnetic simulation software Mumax3 used in this work is open-source and can be accessed freely at http://mumax.github.io/.

\medskip
\textbf{Conflict of Interest} \par
The authors declare no conflict of interest.

\medskip

%







\end{document}